# When categorization-based stranger avoidance explains the uncanny valley: A comment on MacDorman & Chattopadhyay (2016)

Takahiro Kawabe [a], Kyoshiro Sasaki [b,d], Keiko Ihaya [c], & Yuki Yamada [b]

[a]NTT Communication Science Laboratories, Nippon Telegraph and Telephone Corporation, Japan; [b]Faculty of Arts and Science, Kyushu University, Japan; [c]Admission Center, Kyushu University, Japan; [d]Japan Society for the Promotion of Science, Japan

## Abstract

Artificial objects often subjectively look eerie when their appearance to some extent resembles a human, which is known as the uncanny valley phenomenon. From a cognitive psychology perspective, several explanations of the phenomenon have been put forth, two of which are object categorization and realism inconsistency. Recently, MacDorman and Chattopadhyay (2016) reported experimental data as evidence in support of the latter. In our estimation, however, their results are still consistent with categorization-based stranger avoidance. In this Discussions paper, we try to describe why categorization-based stranger avoidance remains a viable explanation, despite the evidence of MacDorman and Chattopadhyay, and how it offers a more inclusive explanation of the impression of eeriness in the uncanny valley phenomenon.

Correspondence to:
Takahiro Kawabe,
NTT Communication Science Laboratories,
Nippon Telegraph and Telephone Corporation,
3-1, Morinosatowakamiya,
Atsugi, 243-0198,
Kanagawa, Japan

E-mail: kawabe.takahiro@lab.ntt.co.jp
Telephone: +81-46-240-3592
Facsimile: +81-46-240-4716

## Motivation of this commentary

Recently, MacDorman and Chattopadhyay (2016) proposed that the consistency of realism among human morphological features, rather than object categorization, is a critical factor determining eeriness in the uncanny valley phenomenon (Mori, MacDorman, & Kageki, 2012). Although we understand what they are suggesting, at this stage we feel that the critical part of their proposal is also compatible with the categorization-based explanation, that is, with categorization-based stranger avoidance (Yamada, Kawabe, & Ihaya, 2013). Here, we describe why we believe this and discuss scenarios where the categorization-based stranger avoidance can still be one of the factors contributing to the production of the uncanny valley phenomenon.

## Categorization-based stranger avoidance

First, let us describe how categorization-based stranger avoidance can explain eeriness in the uncanny valley phenomenon. Categorization-based stranger avoidance was first proposed by Yamada, Kawabe, and Ihaya (2013). In their Experiment 1, they used morphed images of real, stuffed, and cartoon human faces as stimuli and measured both a likability score and categorization latency to classify each image into a given categorization class. They found that the likability score was lowest when the categorization latency was highest. In their Experiment 2, they replicated the results using stimulus images of real, stuffed, and cartoon dog faces.

Importantly, in their Experiment 3, they observed that the likability score did not drop even when the categorization latency increased, wherein the categories the observers needed to judge were both related to humans (gender or identity).

The results of Experiments 1 and 2 are apparently inconsistent with the results of Experiment 3; however all of the results can be consistently interpreted by the categorization-based stranger avoidance Yamada, Kawabe, and Ihaya (2013) proposed. In Experiments 1 and 2, images with the highest categorization latency could have an improbable appearance, such as a cartoon-like real human face. In this case, the observers might first try to categorize the stimuli into a novel class. In other words, an object with such an improbable appearance is not categorized into already acquired classes of objects, is probably judged as a stranger to be avoided, and, consequently, has a low likability score. In this scenario, the categorization latency might increase because the object category inferred from the stimulus appearance does not exist in a given set of category classes for the task. On the other hand, in Experiment 3, images with the highest categorization latency could have a probable appearance, such as a boyish girl face. In this case, the observer might try to categorize the stimuli into a familiar class. That is, an object with such a probable appearance is categorized into already acquired classes of objects, is not judged to be a stranger to be avoided, and thus does not have low likability. In this scenario, the object category exists in a given set of categorization classes. On the other hand, because the category classes are too similar to each other, it may take time for the observer to discern which category is appropriate for the task.

In this respect, we emphasize that stranger avoidance is not driven simply by the categorization difficulty that can be quantified by measuring categorization latency. Rather, it is triggered when an object has an improbable appearance and is therefore categorized into a novel class.

The existence of categorization-based stranger avoidance is supported by other lines of research. Using morphed images of a tomato and a strawberry as stimuli, Yamada, Kawabe, and Ihaya (2012) investigated how observers judged the willingness to eat. Consistent with Yamada, Kawabe, and Ihaya (2013), the results showed that the rating score for the willingness to eat was lowest when the rating scores for categorization confidence was lowest and thus when the food category was perhaps improbable (such as a tomato-like strawberry). In addition, they investigated how individual differences in food avoidance behavior, i.e., food neophobia (Rozin & Fallon, 1980), could modulate the willingness to eat and found that the rating score for the willingness to eat further decreased for participant groups with high food neophobia scores when categorization confidence was lowest. That the negative impression of foods in morphed images was enhanced by individual differences in food neophobia indicates that the categorization of the foods in morphed images into a novel class possibly causes the negative impression of the foods, and it supports the idea that the difficulty in categorizing an object into a novel class is one of decisive factors determining the eeriness impression.

### How stranger avoidance explains MacDorman and Chattopadhyay

MacDorman and Chattopadhyay (2016) used morphing images of a 3D model face and real human face and investigated the relationship between the categorization difficulty (or categorization ambiguity) and eeriness impressions. They found that the eeriness impression did not peak at the morph rate with the most ambiguous categorization and that it basically decreased as the morph rate of the real human face with the 3D model face increased while horizontal shifts of eeriness impression functions occurred in the morph rate dimension. On the basis of the data, MacDorman and Chattopadhyay tried to rule out categorization-related mechanisms from the explanation for eeriness in the uncanny valley phenomenon.

On the other hand, we suggest that their results are not sufficient to deny the involvement of object categorization in the uncanny valley phenomenon and that they are indeed consistent with the categorization-based stranger avoidance explanation. As MacDorman and Chattopadhyay acknowledged, the 3D model was the eeriest among the stimuli they used. That is, it could have the most improbable appearance of "a real human who has 3D-model like facial features." The stranger avoidance theory predicts that the eeriness impression will decrease as the morph rate increases in the stimulus setting of MacDorman and Chattopadhyay because their stimulus images gradually take on a more probable appearance of "a real human who has real human-like facial features." As the morphing rate increases, and the image can be better categorized into a familiar class. Actually, this was what MacDorman and Chattopadhyay observed. Cheetham, Suter, and Jancke (2014) also reported non-negative impressions for an ambiguous figure created by morphing between avatar and human images. The results are also consistent with categorization-based stranger avoidance: because the avatar originally produces the most negative impressions, the morphed image between it and human faces cancels the avatar's improbable appearance and the image is more likely categorized into a familiar class such as human.

MacDorman and Chattopadhyay explained why they used the limited range of stimulus category in the following way: "From the standpoint of experimental control, it is difficult to investigate transitions along a human similarity dimension" (page 192). We appreciate that this is, in a sense, the proper attitude for a scientist who is trying to appropriately manipulate experimental components without potential artifacts. However, to rule out the involvement of the categorization-based explanation in a fair manner, they should have used a stimulus category dimension with a wider range containing a non-human entity, an ambiguous entity, and a human being. Although MacDorman and Chattopadhyay pointed out the potential involvement of morphing artifacts in previous studies, not all of the morphing images used as stimuli in those studies necessarily had such artifacts that caused negative impressions. Moreover, recent studies (Ferrey, Burleigh, & Fenske, 2015; Sasaki, Ihaya, & Yamada, submitted) show that the eeriness impression is strongest when categorization is the most ambiguous and possibly the most improbable. In addition, in Ferrey, Burleigh, and Fenske (2015), the eeriest impressions occurred when they used morphed images between different kinds of animals as stimuli. The eeriness impression's occurring between different animal categories is out of scope for the realism inconsistency hypothesis but is well explained by the categorization-based stranger avoidance.

### Two theories: Independent of each other?

We have so far argued that the results reported by MacDorman and Chattopadhyay are also consistent with the categorization-based stranger avoidance hypothesis. We do not intend to imply that our theory is an exclusive explanation of the eeriness in the uncanny valley phenomenon. Actually, our theory relies on the perception of object appearance improbability to determine how novel an object category is. The evaluation of the appearance improbability is consistent with the evaluation of the realism inconsistency (MacDorman & Chattopadhyay, 2016) or perceptual mismatch (Brenton, Gillies, Ballin, & Chatting, 2005; MacDorman, Green, Ho, & Koch, 2009; Kätsyri, Förger, Mäkäräinen, & Takala, 2015). On the other hand, our theory posits that after recognizing the improbable appearance, the brain categorizes the object into a novel class and recognizes the object as a stranger to be avoided. We consider categorization-based stranger avoidance to be a kind of hybrid idea between category-based explanations and appearance-based explanations such as realism inconsistency and perceptual mismatch. Our definitions of object categorization and improbable appearance are not new and are possibly similar to the definitions (i.e., category conflict and feature atypicality) proposed in a previous study (Burleigh, Schoenherr, & Lacroix, 2013).

The uncanny valley phenomenon has been found in a continuum of stimulus appearances from non-human to human (Mori, MacDorman, & Kageki, 2012), but there is room to discuss whether it can be extended to dimensions not directly related to human. The appearance-based explanation emphasizes the deviation of an object's appearance from a human one as the decisive factor causing eeriness in the uncanny valley phenomenon (Brenton, Gillies, Ballin, & Chatting, 2005; MacDorman, Green, Ho, & Koch, 2009; Kätsyri, Förger, Mäkäräinen, & Takala, 2015). It thus focuses on human likeness in the stimulus continuum from non-human to human. On the other hand, the categorization-based explanation does not always focus on human likeness. It thus assumes a more general cognitive function to emotionally evaluate an object on the basis of categorization (Yamada, Kawabe, & Ihaya, 2012, 2013, Yamada, Sasaki, Kunieda, & Wada, 2014). Unfortunately, no study has thoroughly investigated whether the mechanism underlying eeriness impressions is different between the continuum between non-human and human categories and the continuum between two non-human categories. Addressing this issue may provide more suitable explanations for the uncanny valley phenomenon.

### Relation with inhibitory devaluation

The eeriness impression can be explained in terms of an idea that is different from our theory but based on object categorization. One of the strongest theories is inhibitory devaluation (Ferrey, Burleigh, & Fenske, 2015). Ferrey, Burleigh, and Fenske (2015) reported experimental data similar to the data of Yamada, Kawabe, and Ihaya (2013) but gave a different account of the data. In their theory, when an object categorization is ambiguous, inhibition is triggered to resolve conflict between competing stimulus-related representations, and this causes the inhibitory devaluation. The inhibitory devaluation is an emotional evaluation phenomenon that generally occurs in attentionally inhibited object (Frischen, Ferrey, Burt, Pistchik, & Fenske, 2012; Raymond, Fenske, & Tavassoli, 2003).

We acknowledge that inhibitory devaluation can explain most of the reported data on the eeriness impression in the uncanny valley phenomena. On the other hand, it does not fully explain the data reported in Experiment 3 of Yamada, Kawabe, and Ihaya, wherein they showed that the morphed image between a male and female facial photograph (or between different male individuals) did not strongly produce eeriness. It is possible that because the morphed images between a male and female are categorized into either of familiar classes (male or female), the devaluation might not occur. For the theories of both stranger avoidance and inhibitory devaluation, there are issues to address. For stranger avoidance, it is necessary to further test when the difficulty in categorizing an object into a novel class occurs. For inhibitory devaluation, it is necessary to examine what category condition causes the devaluation.

**Acknowledgments**

This study was supported by JSPS KAKENHI Grant Number JP14J06025 given to K.S., JSPS KAKENHI Grant Number JP26540067 and JP15H05709 given to Y.Y., JSPS KAKENHI Grant Number JP26750322 given to K.I., and Kyushu University Interdisciplinary Programs in Education and Projects in Research Development (#27822) given to Y.Y.